\documentclass{sig-alternate-05-2015}

\usepackage[utf8]{inputenc} 
\usepackage[T1]{fontenc}
\usepackage{microtype} 

\usepackage{balance}
\usepackage{cite}
\usepackage[vlined,figure,linesnumbered]{algorithm2e}
\usepackage{booktabs}
\usepackage{color}
\usepackage{amsmath}
\usepackage[table]{xcolor}
\usepackage{siunitx}
\usepackage{hyperref}
\usepackage{pgfplots}
\pgfplotsset{compat=newest}

\definecolor{dkgreen}{rgb}{0,0.5,0}
\definecolor{dkred}{rgb}{0.5,0,0}
\definecolor{dkgray}{rgb}{0.3,0.3,0.3}
\usepackage{tikz}
\usetikzlibrary{arrows,shadows,positioning,shapes}

\makeatletter
\providecommand*{\cupdot}{%
  \mathbin{%
    \mathpalette\@cupdot{}%
  }%
}

\newcommand*{\@cupdot}[2]{%
  \ooalign{%
    $\m@th#1\sqcup$\cr
    \hidewidth$\m@th#1\cdot$\hidewidth
  }%
}
\makeatother

\usepackage{listings}
\newcommand{\lt}[1]{{\lstinline@#1@}}

\newcommand{\aset}[1]{\{#1\}}
\newcommand{\punion}{\cupdot}
\newcommand{\hdrc}[1]{\multicolumn{1}{c}{\cellcolor{black!30}{\textsf{#1}}}}
\newcommand{\hdrr}[1]{\multicolumn{1}{r}{\cellcolor{black!30}{\textsf{#1}}}}

\newcommand{\hdrv}[1]{\cellcolor{black!30}{\textsf{#1}}}
\newcommand{\hdrt}[1]{\textsf{#1}}
\newcommand{\pfmt}[1]{\textsf{#1}}

\newboolean{long}
\font\xtiny=cmr10 scaled 600
\definecolor{siqrcolor}{gray}{0}

\def\tableentryraw#1#2{\num{#1}\parbox{16pt}{\raggedleft\color{siqrcolor}\xtiny#2\normalcolor}}

\def\mso#1#2{
	\ifx\\#1\\%
		-
	\else%
		\ifthenelse{\equal{#1}{0}}{\tableentryraw{#1}{#2}}{
			\ifthenelse{\boolean{long}}{\tableentryraw{#1}{#2}}{#1}}
	\fi%
}

\begin{document}

\newif\ifcopyright
\copyrightfalse

\ifcopyright
\setcopyright{acmcopyright}
\else
\setcopyright{none}
\fi

\CopyrightYear{2016} 
\conferenceinfo{FSE'16,}{November 13-19, 2016, Seattle, WA, USA}
\isbn{978-1-4503-4218-6/16/11}\acmPrice{\$15.00}
\doi{http://dx.doi.org/10.1145/2950290.2950311}

\title{iGen: Dynamic Interaction Inference for Configurable Software}

\numberofauthors{1}
\author{
  \alignauthor ThanhVu Nguyen \hspace*{0.8em} Ugur Koc  \hspace*{0.8em}
  Javran Cheng  \hspace*{0.8em} Jeffrey
  S. Foster  \hspace*{0.8em} Adam A. Porter
  \affaddr{University of Maryland, College Park, USA}\\
  \email{\{tnguyen, ukoc, javran, jfoster, aporter\}@cs.umd.edu}
}

\maketitle

\begin{abstract}
  To develop, analyze, and evolve today's highly configurable software
  systems, developers need deep knowledge of a system's configuration
  options, e.g., how options need to be set to reach certain
  locations, what configurations to use for testing, etc. Today,
  acquiring this detailed information requires manual effort that is
  difficult, expensive, and error prone. In this paper, we propose
  iGen, a novel, lightweight dynamic analysis technique that
  automatically discovers a program's \emph{interactions}---expressive
  logical formulae that give developers rich and detailed information
  about how a system's configuration option settings map to particular
  code coverage. iGen employs an iterative algorithm that runs a
  system under a small set of configurations, capturing coverage data;
  processes the coverage data to infer potential interactions; and
  then generates new configurations to further refine interactions in
  the next iteration. We evaluated iGen on 29 programs spanning
  five languages; the breadth of this study would be
  unachievable using prior interaction inference tools. Our results
  show that iGen finds precise interactions based on a very small
  fraction of the number of possible configurations. Moreover, iGen's
  results confirm several earlier hypotheses about typical interaction
  distributions and structures.
\end{abstract}



\section{Introduction}
\label{sec:intro}

Modern software systems are increasingly designed to be
configurable. This has many benefits, but it also greatly complicates
tasks such as testing, debugging, and impact analysis because the
total number of configurations can be very large. To carry out such
tasks, developers must take advantage of task-specific structure in a
system's configuration space. For example, a developer may observe that
many configurations are the same in terms of the coverage they achieve
under a test suite, and thus developers can perform effective testing
using just a small set of configurations.

In prior work~\cite{reisner2010using, song:icse12, song2014itree}, we
showed how to automatically infer \emph{interactions} that concisely
describe a system's configurations.  Our focus is coverage, and we
formally define an \emph{interaction} to be a formula $\phi$ over
configuration options such that (a) any configuration satisfying $\phi$
covers some location $L$ under a given test suite and (b) $\phi$ is the logically weakest such
formula (i.e., if $\psi$ also describes configurations covering $L$
then $\psi \Rightarrow \phi$). Thus, by knowing a system's
interactions, a developer can determine useful information
about configurations, e.g., given a location, determine what configurations cover
it; given an interaction, determine what locations it covers; find important options and compute a minimal set of configurations to achieve certain coverage; etc.
In the literature, feature interactions and presence conditions (Section~\ref{sec:related}) are similar to interactions and can explain functional (e.g., bug triggers, memory leaks) and non-functional (e.g., performance anomalies, power consumption)
behaviors. Interactions can also aid reverse engineering and impact
analysis~\cite{She:2011:REF:1985793.1985856,Berger:2010:VMR:1858996.1859010}.

While our prior work was promising, it has significant limitations. In
our first effort, we inferred interactions using Otter, a symbolic
executor for C~\cite{reisner2010using}. However, symbolic execution
does not scale to large systems, even when restricted to configuration options; is brittle in the presence of frameworks, libraries, and native code; and is
language-specific. In our second effort, we developed
iTree~\cite{song2014itree}, which infers interactions using dynamic
analysis and machine learning. However, iTree's focus is on finding
configurations to maximize coverage, and in practice it only
discovers a small set of interactions. (Section~\ref{sec:related}
discusses these prior systems in more detail.)

In this paper, we introduce iGen, a new dynamic analysis tool for
automatically discovering a program's interactions. iGen works by
iteratively running a subject program under a test suite and set
of configurations; inferring potential interactions from the resulting
coverage information; and then generating new configurations that aim to
refine the inferred interactions on the next iteration. By carefully choosing
new configurations in this last step, iGen is able to quickly converge
to a final, precise set of interactions using only a small set of
configurations.

Moreover, iGen's design overcomes the limitations of Otter and
iTree. The only language-specific portion of iGen is obtaining code
coverage, which is widely available for almost any language. iGen's
analysis is also very lightweight and scalable, because, as we explain
later, it uses simple computations over coverage information to
infer interactions. Finally, although we did not mention it earlier,
our prior work was restricted to inferring interactions that are
purely conjunctive. In contrast, iGen supports interactions that are
purely conjunctive, purely disjunctive, and specific mixtures of the
two. (Section~\ref{sec:algorithm} describes iGen in more detail.)

We evaluated iGen by running it on 29 programs, far more than
we were ever able to use with Otter or iTree. Our subject programs
span five languages (C, Perl, Python, Haskell, and OCaml); range in
size from tens of lines to hundreds of thousands of lines; and have
between 2 and 50 configuration options. For most programs we
apply iGen to run-time configuration options, but for one program,
\pfmt{httpd}, we study compile-time configuration
options. (Section~\ref{sec:benchmarks} describes our subject programs.)

We considered three research questions. First, we evaluated the
correctness of iGen's inferred interactions. We found that, for a
subset of the subject programs, the interactions produced by iGen's
iterative algorithm are largely similar to what iGen would produce if
it inferred interactions from \emph{all} possible configurations. This
suggests iGen does converge to the optimal solution. We also manually
inspected a subset of the interactions found by iGen and verified they
match the logic in the code.

Second, we measured iGen's performance and found it explores a
very small fraction of the number of possible configurations. 
Moreover, like the work of Reisner et al., iGen generates dramatically
more precise interactions than iTree. Yet, it runs in a small fraction
of the time required for Reisner et al.'s experiments.

Finally, we analyzed iGen's output to learn interesting properties
about the subject programs. We confirmed several results found in
prior work~\cite{reisner2010using,song:icse12,song2014itree}, among
them: the number of interactions is far smaller than what is
combinatorially possible; yet a few (very) long interactions are needed
for full coverage; and \emph{enabling options}, which must be set a
certain way to achieve most coverage, are common. 
 We should emphasize that our prior
work hypothesized these based cumulatively on just four programs in
one language, whereas we observe them based on 29 programs in
five languages. We also observed new phenomena: that
disjunctive and mixed interactions are less common than conjunctive ones, but
nonetheless cover a non-trivial number of lines. 
Finally, we showed that iGen's interactions can be used to
compute a small set of configurations that cover all or most lines of
the programs.
(Section~\ref{sec:eval} reports on our evaluation.)

We believe iGen takes an important step forward in the practical
understanding of configurable systems.

\section{iGen Algorithm}
\label{sec:algorithm}

\begin{algorithm}[t]
\footnotesize
\DontPrintSemicolon

\SetKwInOut{Input}{input}
\SetKwInOut{Output}{output}
\SetKwFunction{oneWayCoveringArray}{oneWayCoveringArray}
\SetKwData{defaultconfig}{default\_config}
\SetKwData{configs}{configs}
\SetKwData{cov}{cov}
\SetKwData{ncov}{ncov}
\SetKwData{oldcov}{old\_cov}
\SetKwData{subncov}{subncov}
\SetKwData{subcov}{subcov}
\SetKwData{inter}{ints}
\SetKwData{oldinter}{old\_ints}
\SetKwFunction{runTestSuite}{runTestSuite}
\SetKwFunction{break}{break}
\SetKwData{selinter}{sel\_inter}
\SetKwFunction{genNewConfigs}{genNewConfigs}
\SetKwFunction{select}{select}
\SetKwData{cover}{cover}
\SetKwData{noncover}{noncover}
\SetKwFunction{cond}{cond}
\SetKwFunction{ncond}{ncovcond}
\SetKwFunction{conj}{conj}
\SetKwFunction{disj}{disj}
\SetKwFunction{conjm}{conjdisj}
\SetKwFunction{disjm}{disjconj}
\SetKwFunction{check}{check}
\SetKwFunction{selStrongest}{selStrongest}
\SetKwFunction{result}{result}

\Input{a program $P$ and a test suite $T$} 
\Output{a set of interactions of $P$}
\BlankLine

$\cov, \; \inter \leftarrow \emptyset$\;
$\configs \leftarrow \oneWayCoveringArray{} \cup \{\defaultconfig\}$\; \label{line:init}
\BlankLine

\While{$true$}{
  \oldcov $\leftarrow$ \cov\;
  \oldinter $\leftarrow$ \inter\;
  $\cov, \ncov \leftarrow \runTestSuite(P,T,\configs)$\; \label{line:testsuite}
  // returns $\cov(l) = \{ c \mid c \textrm{ covers } l \}$\;
  // \phantom{returns} $\ncov(l) = \{ c \mid c \textrm{ does not cover } l \}$

  \BlankLine
  \ForEach {location $l \in \cov$}{ \label{line:loopstart}

    $\conj \leftarrow \punion \; \cov(l)$\; \label{line:conj}
    \BlankLine

    $\disj \leftarrow \neg (\punion \; \ncov(l))$\; \label{line:disj} 

    \BlankLine

    $\disj' \leftarrow \neg \punion \{c \mid c \in \ncov(l) ~\wedge~ c \Rightarrow \conj\}$\; \label{line:conjmstart}
    $\conjm \leftarrow \conj \wedge \disj'$\; \label{line:conjmend}

    \BlankLine

    $\conj' \leftarrow \punion \{c \mid c \in \cov(l) ~\wedge~ \disj \Rightarrow \neg c\}$\; \label{line:disjmstart}
    $\disjm \leftarrow \disj ~\vee~ \conj'$\; \label{line:disjmend}

    \BlankLine
    $\inter(l)\leftarrow(\conj, \disj, \conjm, \disjm)$\; \label{line:setints}
  }
  \BlankLine

  \lIf{$\cov = \oldcov \wedge \inter = \oldinter$}{
    \break
  } \label{line:break}

  \BlankLine
  \configs $\leftarrow$ \genNewConfigs{\inter}\; \label{line:gennewconfigs}

}

\BlankLine
\ForEach {location $l \in \cov$}{ \label{line:postprocess}
  $\inter(l) \leftarrow \check(\inter{l}, \cov(l))$\; \label{line:check}
  $\result(l) \leftarrow \selStrongest(\inter{l})$\; \label{line:selstrongest}
}
\BlankLine

\KwRet{\result} 
\caption{iGen's iterative algorithm for inferring program interactions.}
\label{alg:igen}
\end{algorithm}

We begin our presentation by describing the iGen algorithm, 
whose pseudo-code is shown in Figure~\ref{alg:igen}. 
The input to iGen is a program $P$ and a test suite $T$, 
and the output is a set of interactions for locations in $P$ 
that were covered when running on $T$. 
iGen works by iteratively generating 
a set of configurations (\textsf{configs} in the algorithm) 
until the coverage (\textsf{cov}) and interactions (\textsf{ints}) 
inferred from that set reach a fix-point.

The algorithm begins on line~\ref{line:init} by initializing \textsf{configs} to a randomly generated 1-way covering array~\cite{cohen2003constructing,cohen1996combinatorial}, i.e., it contains all possible settings of each individual option.
The algorithm also includes a default configuration if one is
available. In our experience, such a configuration typically yields
high coverage under the test suite and hence is a useful starting
point.
On line~\ref{line:testsuite}, iGen runs the test suite under the current set of configurations,\footnote{In practice iGen memoizes the  coverage information from previous runs and only runs the test suite  under the new configurations.} producing two coverage maps: \textsf{cov} maps each location $l$ to the set of configurations $c$ such that at least one test covers $l$ under $c$, and  \textsf{ncov} maps $l$ to the set of configurations that do not cover $l$.

Then for each location $l$ covered by the test suite under some configuration (line~\ref{line:loopstart}), iGen infers candidate interactions. 
Although in theory interactions can be arbitrary formulae, iGen keeps its inference process efficient by assuming interactions follow particular syntactic \emph{templates}. 
As it iterates, iGen computes the most precise interaction for each location for each template. 
At the end of the algorithm, iGen selects the \emph{strongest} (in a logical sense) interaction per location across the different templates.

Currently, iGen supports four templates: $\textsf{conj}$, a purely
conjunctive interaction; $\textsf{disj}$, a purely
disjunctive interaction; $\textsf{conjdisj}$, a conjunctive
interaction where the last conjunct is a disjunct; and
$\textsf{disjconj}$, a disjunctive interaction where the last
disjunct is a conjunct. We explain the computation of the interactions
in detail below.  This particular set of
templates was chosen partially based on our experience (e.g., we
believe conjunctive interactions are very common) and partially based
on what is efficient to compute (e.g., mixing one disjunction into a
conjunction or vice-versa is a relatively small cost, whereas more
complex interleavings of conjunctions and disjunctions would be much
less efficient.)

After saving candidate interactions (line~\ref{line:setints}), the loop
terminates if iGen has reached a fix-point (line~\ref{line:break}). Otherwise, iGen
creates additional configurations (line~\ref{line:gennewconfigs}) designed to refine
interactions (details below) and continues iteration.

After the main loop terminates, there are two steps remaining. First,
because of some heuristics in iGen's interaction generation,
some interactions it computes may not actually cover the expected
lines. Thus on line~\ref{line:check}, iGen iterates through the set of
interactions and checks that for any interaction $\phi$ for $l$,
it is actually the case that $c \Rightarrow \phi$ for all configurations $c$
that cover $l$. iGen eliminates any interaction that fails this check
by setting it to \emph{true}.
Second, on line~\ref{line:selstrongest}, iGen sets
$\textsf{result}(l)$ to be the logically strongest interaction among
 $\textsf{conj}$, $\textsf{disj}$, $\textsf{conjdisjm}$, and $\textsf{disjconj}$. 
If there is no single strongest
interaction, iGen eliminates any interactions that are weaker than
another and returns the conjunction of the remaining strongest
interactions.

\paragraph*{Running Example}

\begin{figure}[t]
 \begin{lstlisting}[numbers=none,xleftmargin=1em,language=C]
// options: $s$, $t$, $u$, $v$, $x$, $y$, $z$
int max_z = 3;

if($x$ && $y$) {
  printf("$L0$\n"); // $x \wedge y$
  if (!(0 < $z$ && $z$ < max_z)){
    printf("$L1$\n"); // $x \wedge y \wedge (z\in\aset{0, 3, 4})$
  }
}else{
  printf("$L2$\n"); // $\neg{x} \vee \neg{y}$
}
printf("$L3$\n");  // $\textit{true}$
if($u$ && $v$) {
  printf("$L4$\n");  // $u \wedge v $
  if($s$ || $t$) {
    printf("$L5$\n");  // $u \wedge v \wedge (s \vee t) $
  }
}
\end{lstlisting}
\caption{Program with seven configuration options. 
Locations L0--L5 are annotated with associated interactions.}
\label{fig1}
\end{figure} 

We next use the C program in Figure~\ref{fig1} to explain the details of iGen.
This program has seven configuration options, listed on the first line of the figure. 
The first six options are boolean-valued, and the last one, $z$, ranges over the set $\{0,1,2,3,4\}$.
Thus, this program has $2^6 \times 5 = 320$ possible configurations.

The code in Figure~\ref{fig1} includes print statements that mark six locations $L0$--$L5$.
At each location, we list the associated desired interaction.
For example, $L1$ is covered by any configuration in which $x$ and $y$ are true and $z$ is 0, 3, or 4.
As another example, $L3$ is covered by any configuration, hence its interaction is \textit{true}.

Prior approaches to interaction inference are not sufficient for this
example. The work of Reisner et. al~\cite{reisner2010using}
only supports conjunctions, so it must approximate the interactions for $L1$, $L2$, and $L5$.
iTree \cite{song2014itree}
actually produces no interactions for this example, because all lines
are covered by iTree's initial two-way covering array (iTree stops
generating interactions when no new coverage is achieved).

For this example, iGen initializes \textsf{configs} to the following
covering array (there is no default configuration):
\[
\footnotesize
\begin{array}{c|rrrrrrr|l}
\text{config}& s & t & u & v & x & y & z & \text{coverage}\\
\midrule
c_1 & 0 & 0 & 1 & 1 & 1 & 0 & 1 & L2,L3,L4 \\
c_2 & 1 & 1 & 0 & 0 & 1 & 1 & 0 & L0,L1,L3\\
c_3 & 0 & 0 & 1 & 1 & 0 & 0 & 2 & L2,L3,L4 \\
c_4 & 0 & 0 & 1 & 1 & 1 & 1 & 3 & L0,L1,L3,L4 \\
c_5 & 0 & 1 & 1 & 1 & 1 & 0 & 4 & L2,L3,L4,L5  \\
\end{array}
\]
We list the coverage of each configuration on the right.

\paragraph*{Conjunctive Interactions}
The first interaction template, $\textsf{conj}$, supports
conjunctions of \emph{membership constraints} $x\in S$ indicating
option $x$ ranges over set $S$.  For example, the interaction for $L1$
in Figure~\ref{fig1} is shorthand for
$(x\in\aset{1})\wedge (y\in\aset{1})~\wedge~ (z\in\aset{0,3,4})$. On
line~\ref{line:conj} of the algorithm, iGen infers $\textsf{conj}$ by taking the
\emph{pointwise union} of the covering configurations' option settings
and then conjoining them. We denote this operation by $\punion$. For
example, the table below shows the pointwise union of the two covering
configurations $c_2$ and $c_4$. Here $\top$ is the universal set for
an option.
\[
\footnotesize
\begin{array}{r|rrrrrrr}
L1 & s & t & u & v & x & y & z \\
\midrule
c_2   & 1 & 1 & 0 & 0 & 1 & 1 & 0\\
c_4   & 0 & 0 & 1 & 1 & 1 & 1 & 3\\
\midrule
\textrm{union} & \top & \top & \top & \top & 1 &1 & 0, 3 \\
\end{array}
\]
Thus to form $\textsf{conj} = c_2 ~\punion~ c_4$ for $L1$ we simply conjoin
the option settings from the above table to yield
$\textsf{conj} = x \wedge y \wedge (z\in\aset{0,3})$, where we write $x$
and $y$ for $x\in\aset{1}$ and $y\in\aset{1}$. Note we omit
constraints corresponding to $\top$, since those indicate options that
can take any value.

At this point, $\textsf{conj}$ is close to, but not quite, the
correct interaction for $L1$. The problem is that \textsf{configs}
is missing a configuration where $x=y=1$ and $z=4$.
Thus---skipping over the other templates and other locations for the
moment---for the next iteration iGen generates additional 
configurations to \emph{refine} the set of interactions, using the
\textsf{genNewConfigs} call on line~18.

iGen derives these new configurations by systematically changing the
settings from one selected interaction from \textsf{ints}.
For example, \textsf{genNewConfigs} might generate new configurations
from interaction $\textsf{conj}$ to yield:
\[
\footnotesize
\begin{array}{c|rrrrrrr|l}
\text{config}& s & t & u & v & x & y & z & \text{coverage}\\
\midrule
c_6    & 1 & 0 & 1 & 0 & 0 & 1 & 0 & L2,L3\\
c_7    & 0 & 0 & 0 & 1 & 1 & 0 & 3 & L2,L3\\
c_8    & 1 & 1 & 0 & 1 & 1 & 1 & 1 & L0,L3\\
c_9    & 1 & 0 & 1 & 0 & 1 & 1 & 2 & L0,L3\\
c_{10} & 1 & 0 & 0 & 1 & 1 & 1 & 4 &  L0,L1,L3
\end{array}
\]
Here each configuration disagrees with $\textsf{conj} = x \wedge y \wedge (z\in\aset{0,3})$ in one
setting, e.g., $c_6$ has $\neg x$, $c_7$ has $\neg y$, and $c_8$ has $z=1$.
Then the next iteration of the fix-point loop will compute
$\textsf{conj}$ for $L1$ from $c_2$, $c_4$, and $c_{10}$. 
Since $c_{10}$
has $x=y=1$ and $z=4$ (which was not covered in the first set of
configurations), iGen produces the correct
interaction $x \wedge y \wedge (z\in \aset{0,3, 4})$.

In practice, we could choose any interaction for any line and use it
to generate new configurations. Currently, iGen's heuristic is to
choose the \emph{longest} current interaction, based on our prior
experience suggesting long interactions are uncommon and hence likely
to be inaccurate. If there is a tie for longest interaction, iGen
selects randomly among the longest. 
If iGen selects an interaction
that does not fully constrain some configuration options, then it 
assigns random values (satisfying whatever constraint in present)
to those options when creating new configurations.

\paragraph*{Disjunctive Interactions}
Next let us consider the interaction $\neg{x} ~\vee~ \neg{y}$ for $L2$
in Figure~\ref{fig1}. By construction, $\textsf{conj}$ cannot encode
this formula---although the membership constraints in $\textsf{conj}$
are a form of disjunction (e.g., $z\in\aset{0,3}$ is the same as
$z=0 \vee z=3$), they cannot represent disjunctions among different
variables.

There are a variety of potential ways to infer more general
disjunctions, but we want to maintain the same efficiency as inferring
conjunctive interactions. To motivate iGen's approach to disjunctions,
observe that $L2$'s interaction arises because an else branch was
taken. In fact, $L2$'s interaction is exactly the negation of the
interaction for $L0$ from the true branch. Thus,
iGen computes disjunctive interactions by first computing a non-covering
interaction, which is a \emph{conjunctive} interaction
for the configurations that do \emph{not} cover line $L2$,
and then \emph{negates} it to get a disjunctive
interaction for $L2$ (line~\ref{line:disj}).
In our running example, $c_2$ and $c_4$ are the only
configurations that do not cover $L2$, thus iGen computes 
$c_2\punion c_4 = x\wedge y\wedge (z\in\aset{0,3})$. 
Negating that yields $\textsf{disj} =\neg{x} \vee \neg{y} ~\vee~ (z\in\aset{1,2,4})$, which is 
close to the correct interaction for $L2$.

Notice this approach to
disjunctions is a straightforward extension of conjunctive interaction
inference. Also notice that it is heuristic since the computed
interaction may not actually cover the given line; thus disjunctive interactions
may be eliminated on line~\ref{line:check} of the algorithm in Figure~\ref{alg:igen}.

Disjunctive interactions can be refined in two ways. First, they may be
refined by coincidence if \textsf{genNewConfigs} selects a long
conjunctive interaction to refine. 
Second, \textsf{genNewConfigs} also considers the negation of
$\textsf{disj}$ as a possible longest interaction to use for
refinement (essentially refining an interaction describing configurations
that do not reach the current location).

\paragraph*{Mixed Interactions}
Finally, some interactions require mixtures of conjunctions and
disjunctions, such as the interaction $u\wedge v \wedge (s\vee t)$ for
$L5$. Looking at Figure~\ref{fig1}, notice this interaction occurs
because a disjunctive condition is nested inside of a conjunctive
condition---in fact, the interaction for $L5$ is the interaction for
$L4$ with one additional clause.

This motivates iGen's approach to inferring mixed interactions by
\emph{extending} shorter
interactions. Lines~\ref{line:conjmstart}--\ref{line:conjmend} give
the code for computing $\textsf{conjdisj}$.  Recall that to compute
the pure disjunction $\textsf{disj}$, iGen negates the pointwise
union of non-covering configurations. On line~\ref{line:conjmstart} we
use the same idea to compute $\textsf{disj}'$, but instead of
\emph{all} non-covering configurations, we only include the
non-covering configurations that satisfy $\textsf{conj}$.
Essentially we are projecting the iGen algorithm onto just
configurations that satisfy that interaction. Thus, when we infer the
disjunction $\textsf{disj}'$, we conjoin it onto $\textsf{conj}$ to
compute the final mixed interaction.

For our running example, after several iterations $\textsf{conj}$
for $L5$ will be $u\wedge v$ (details not shown). Out of the configurations
that do not cover $L5$, only $c_1$, $c_3$, and $c_4$ also
satisfy $u\wedge v$. Thus $\textsf{disj}'$ will be
$\neg(c_1\punion c_3\punion c_4) = s \vee \ t \vee \neg u \vee \neg v
\vee (z\in\aset{0,4})$.
Thus after some simplification we get
$\textsf{conj} \wedge \textsf{disj}' = u \wedge v \wedge (s
\vee t \vee (z\in\aset{0,4}))$,
which is almost the interaction for $L5$. 
After further iteration, iGen eventually reaches the final, fully precise
interaction for $L5$.

Using the dual of the above approach,
lines~\ref{line:disjmstart}--\ref{line:disjmend} infer another
mixed interaction $\textsf{disjconj}$ by extending the computed
disjunctive interaction $\textsf{disj}$ with a conjunction
$\textsf{conj}'$. Here $\textsf{conj}'$ is generated just like
$\textsf{conj}$, but we only include configurations that disagree in
some setting with some clause of $\textsf{disj}$, since otherwise the configuration is already
included in the left side of the disjunct on line~\ref{line:disjmend}.

Notice that iGen's approach for inferring mixed interactions maintains
the efficiency of computing pure conjunctions and disjunctions. We
could extend the algorithm further to compute conjunctions with nested
disjunctions with nested conjunctions etc., but we have not explored
that yet. 

Lastly, in addition to considering $\textsf{conj}$ and $\textsf{disj}$
as potential longest interactions, \textsf{genNewConfigs} also
considers $\textsf{conj'}$; and (negated) $\textsf{disj'}$. Thus,
mixed interactions may be refined whenever one of those four
components is refined.





\paragraph*{Discussion}

Putting this all together, after running to
completion, iGen produces the same interactions for our example as in the comments in
Figure~\ref{fig1}.  Moreover, iGen finds these interactions by analyzing
just 37 configurations instead of 320 possible configurations.  The
experiments in Section~\ref{sec:eval} show that iGen analyzes an even
smaller fraction of the possible configurations on programs with a
large number of configuration options.

As mentioned above, iGen has several sources of randomness: the
one-way covering array, the interaction used for generating new
configurations, and the values of un- or under-constrained option settings in those new
configurations. Thus, iGen is actually a stochastic algorithm that may
produce slightly different results each time. However, in our
experiments we demonstrate that the variance is reasonable.

Moreover, the computation of each iteration of iGen is straightforward and efficient.
Pointwise union is linear in the number of configurations and options.
Checking the various implications is done with an SMT solver, which is
very efficient in practice. As discussed in Section~\ref{rq2}, iGen's
running time is mostly consumed by running the test suite (line~\ref{line:testsuite}).

Finally, notice that if iGen were to iterate until it had generated
\emph{all} configurations, then it would be guaranteed to produce
correct interactions if they fall under the given templates. For
example, suppose some location $L$ has a purely conjunctive
interaction $x\wedge y$. Then if we consider the set of \emph{all}
configurations that cover $L$, they all satisfy $x\wedge y$; but, the
set has configurations that differ in every possible way for options
that are not $x$ and $y$. Thus, pointwise union of this set will yield
$x$ and $y$ as \emph{true} and every other option as $\top$. Hence iGen must
produce the correct interaction $x\wedge y$.  Similar arguments follow
for the other interaction templates. In Section~\ref{sec:correctness},
we take advantage of this observation to help evaluate the correctness
of iGen's inferred interactions.

\section{Subject Programs}
\label{sec:benchmarks}

\begin{table}[t]
\small
\centering
\caption{Subject programs.}
\label{tab:stats}
\begin{tabular}{@{}l@{\;}lrr@{\;}rr@{}r@{}}
  \hdrv{prog} & \hdrv{lang} & \hdrc{ver} &\hdrc{loc} & \hdrc{opts}  & \hdrc{cspace} & \hdrc{tests}\\

  \pfmt{id}     & C & 8.23  &  332    &    10  &  \num{1024} &      4     \\
  \pfmt{uname}  & C & 8.23  &  281    &    11  &   \num{2048} &      2     \\
  \pfmt{cat}    & C & 8.23  &  496    &    12  &   \num{4096} &     12     \\
  \pfmt{mv}     & C & 8.23  &  375    &    11  &  \num{5120}  &     14     \\
  \pfmt{ln}     & C & 8.23  &  478    &    12  &  \num{10240}  &     14     \\
  \pfmt{date}   & C & 8.23  &  469    &     7  &  \num{17280} &     11     \\
  \pfmt{join}   & C & 8.23  &  892    &    12  &  \num{18432} &     8     \\ 
  \pfmt{sort}   & C & 8.23  &  \num{3348}   &    22  &  \num{6291456} &      9     \\ 
  \pfmt{ls}     & C & 8.23  &  \num{3545}   &    47  &  \num{3.5e14} & 16 \\ 
  \midrule                    
  \pfmt{p-id}     & Perl  & 0.14      &  131 &  8     &\num{256}&      4     \\
  \pfmt{p-uname}  & Perl  & 0.14      &  25  &  6     & \num{64}&      2     \\
  \pfmt{p-cat}  & Perl  & 0.14      &  47  &  7     &\num{128}&     12     \\
  \pfmt{p-ln}  & Perl  & 0.14      &  62  &   2    & \num{4} &     14     \\
  \pfmt{p-date}   & Perl  & 0.14      &  136 &   5    &\num{3360}   &     11     \\
  \pfmt{p-join}   & Perl  & 0.14		 &  178 &  10    &\num{4608}   &     8     \\ 
  \pfmt{p-sort}   & Perl  & 0.14      &  399 &  11    &\num{2048}   &      9     \\
  \pfmt{p-ls}   & Perl  & 0.14      &  403 &  26    & \num{6.7e7} &     16     \\

  \midrule         
 \pfmt{cloc}   & Perl  & 1.62 & \num{8014} & 19 & \num{524288} & 296 \\ 
 \pfmt{ack}    & Perl  & 2.14 & \num{2711} & 32 & \num{4.3e9} & 5 \\ 
 \pfmt{grin}   & Python & 1.2.1 &  628  &  21  & \num{2097152} & 5 \\ 
 \pfmt{pylint} & Python & 1.3.1 &  \num{7837}  &  29  & \num{5.8e10} &  93 \\ 

 \pfmt{hlint} &Haskell& 1.9.21 & \num{3266}  & 12 & \num{8192} & 594 \\
 \pfmt{pandoc} &Haskell& 1.13.2 & \num{24755}  & 22 & \num{4.0e9} & 42 \\ 
 \pfmt{unison} &OCaml  & 2.48.3 & \num{29796}  & 16 & \num{393216} & 5  \\
 \pfmt{bibtex2html} & OCaml  & 1.98 & 9172   &  33  & \num{1.2e9} & 3 \\ 
 \pfmt{gzip}   & C  & 1.6    &  \num{32080}  &  17  & \num{131072} & 10 \\ 
 \pfmt{httpd}  & C  & 2.2.29 &  \num{238345}  &  50  & \num{1.1e+15} & 400 \\
\midrule         
 \pfmt{vsftpd} & C  &2.0.7   &  \num{10482} &   30  &  \num{2.1e9}  &     64     \\ 
 \pfmt{ngircd} & C  & 0.12.0 & \num{13601} &    13  &  \num{29764}  &    141     \\
\bottomrule
\end{tabular}
\end{table}

iGen is implemented in approximately 2,500 lines of Python.
It uses the Z3 SMT solver~\cite{z3ms} to reason about implications.
We computed line coverage using \pfmt{gcov} for C, \pfmt{python-cov}~\cite{python-cov} for Python,
and \pfmt{MDevel::Cover}~\cite{mdevel} for Perl. 
We computed expression coverage using \pfmt{Bisect}~\cite{Bisect} for OCaml and \pfmt{Hpc}~\cite{hpc} for Haskell.

Our experiments were performed on a 2.40GHz Intel Xeon CPU with 16 GB RAM running RedHat
Enterprise Linux 5.11 (64-bit).
The source code for iGen is available at 
\url{https://bitbucket.org/nguyenthanhvuh/igen}. 

\paragraph*{Programs}
Table~\ref{tab:stats} lists our subject programs.
For each program, we list its name, language, version, and lines of
code as measured by \pfmt{SLOCCount}~\cite{sloccount}.  Note that the
line count is typically higher than the number of locations reachable
by the test suite.  We also report the number of configuration options
(\hdrt{opts}) and the total number of possible configurations
(\hdrt{cspace}).  Finally, we list the number of test cases in the
program's test suite.

The first group of programs comes from the widely used GNU
coreutils. These programs are configured via command-line options.  We
selected a subset of coreutils with relatively large
configuration spaces (at least 1024 configurations each).  The second
group comprises coreutils reimplemented in the Perl Power
Tools (PPT) project~\cite{ppt} (excluding \pfmt{mv} which was not
implemented in PPT). These programs are named as the coreutils programs
but with a prefix of \textsf{p-}.

The third group contains an assortment of programs to demonstrate
iGen's wide applicability. Briefly: \pfmt{cloc} is a lines-of-code
counter; \pfmt{ack} and \pfmt{grin} are grep-like programs;
\pfmt{pylint} and \pfmt{hlint} are static checkers for Python and
Haskell, respectively; \pfmt{pandoc} is a document converter;
\pfmt{unison} is a file synchronizer; 
\pfmt{bibtex2html} converts BibTeX files to HTML; 
\pfmt{gzip} is a compression tool; 
and finally,
\pfmt{httpd} is the well-known Apache http server. Cumulatively these
programs span five languages (two programs per languages) and range
from a few thousand to hundreds of thousands of lines.

The last group comprises \pfmt{vsftpd}, a highly secure ftp server,
and \pfmt{ngircd}, an IRC daemon.  These programs were also studied by
Reisner et al.~\cite{reisner2010using}, who used Otter, a symbolic
execution tool, to exhaustively compute all possible program
executions under all possible settings of certain configuration
options.  To make a direct comparison to Reisner et al.'s work
possible, we ran iGen on these programs in a special mode in which,
rather than running a test suite, we used Otter's output as an oracle
of which lines are reachable under which configurations.

\paragraph*{Configuration Options}

We selected configuration options for study in a variety of ways. We
studied all options for coreutils. Most of these options are
boolean-valued, but nine can take on a wider but finite range of values, all of
which we included, e.g., we include all possible formats \pfmt{date} accepts. 
We omit options that range over an unbounded set of values.
For PPT, we studied the same options as coreutils when available,
though PPT only supports a small subset of coreutils' options.

For the programs in the third group, we used the run-time options---for
\pfmt{httpd}, the compile time options---we could get working
correctly. For example, we excluded \pfmt{httpd} options that caused
compiler errors when we changed them.
We ignore options that can take arbitrary values, e.g., \pfmt{pylint}
options that take a regexp or Python expression as input.

Most of the options we selected are
boolean-valued, but several range over a finite set of values, e.g., we
consider seven highlight options for formatting in \pfmt{pandoc} and
three permission modes for modifying files in \pfmt{unison}.

We used the same options for \pfmt{vsftpd} and \pfmt{ngircd} as
Reisner et al., to make a direct comparison possible.

\paragraph*{Test Suites}
We manually created tests for coreutils that cover common
command usage. 
For example, for \pfmt{cat}, we wrote tests that read a text file, a
binary file, a non-existent file, results piped from other commands,
etc.  We used the same tests for coreutils and PPT.

For the third group of programs, test selection varied. For
\pfmt{httpd}, \pfmt{hlint}, \pfmt{pylint}, and \pfmt{bibtex2html}, 
we used the default tests.
For the remaining programs, we started from the
default tests (which were relatively limited) and added more
tests to cover basic functionality.

\section{Evaluation}
\label{sec:eval}
We consider three research questions: 
\begin{itemize}
\item \textbf{R1 (correctness)}: Does iGen generate correct interactions?
\item \textbf{R2 (efficiency)}: What are iGen's performance characteristics?
\item \textbf{R3 (analysis)}: What can we learn from inferred interactions?
\end{itemize}

\setboolean{long}{true}
\begin{table*}[t]
\centering
\caption{iGen's results for the benchmark programs shown in Table~\ref{tab:stats}. Numbers in regular font are medians across 21 runs. Numbers in small font are semi-interquartile ranges measuring variance among the runs.}
\begin{tabular}{lrr|rr|rrrr}
\multicolumn{3}{c}{\cellcolor{black!30}}&\multicolumn{2}{c}{\hdrv time (s)}&\multicolumn{4}{c}{\hdrv interactions}\\
\hdrv{prog} &\hdrc{configs} &\hdrc{cov} & \hdrc{search} &\hdrc{total} &\hdrc{conj} & \hdrc{disj} & \hdrc{mix} & \hdrc{total}\\
\pfmt{id}    & \mso{157}{5}    & \mso{138}{0}   & \mso{18}{1}     & \mso{34}{3}      & \mso{23}{0} & \mso{1}{0} & \mso{1}{0} & \mso{25}{0} \\
\pfmt{uname} & \mso{95}{5}     & \mso{87}{0}    & \mso{9}{1}      & \mso{15}{1}      & \mso{16}{0} & \mso{2}{0} & \mso{7}{1} & \mso{25}{1} \\
\pfmt{cat}   & \mso{131}{6}    & \mso{204}{0}   & \mso{15}{1}     & \mso{42}{5}      &  \mso{18}{0} & \mso{1}{0} & \mso{6}{0} & \mso{25}{0} \\
\pfmt{mv}    & \mso{106}{9}    & \mso{172}{0}   & \mso{9}{1}      & \mso{38}{2}      & \mso{16}{0} & \mso{1}{0} & \mso{1}{0} & \mso{18}{0} \\
\pfmt{ln}    & \mso{213}{18}   & \mso{162}{0}   & \mso{32}{4}     & \mso{96}{13}     & \mso{20}{0} & \mso{1}{0} & \mso{5}{0} & \mso{26}{0} \\
\pfmt{date}  & \mso{680}{44}   & \mso{127}{0}   & \mso{97}{15}    & \mso{350}{94}    & \mso{11}{0} & \mso{1}{0} & \mso{2}{0} & \mso{14}{0} \\
\pfmt{join}  & \mso{323}{21}   & \mso{382}{0}   & \mso{77}{9}     & \mso{158}{25}    & \mso{28}{0} & \mso{6}{0} & \mso{8}{1} & \mso{32}{1} \\
\pfmt{sort}  & \mso{1346}{68}  & \mso{1083}{0}  & \mso{2003}{322} & \mso{3113}{379}  & \mso{78}{1} & \mso{2}{0} & \mso{13}{1} & \mso{93}{2} \\
\pfmt{ls}    & \mso{2175}{250} & \mso{1034}{0}  & \mso{5091}{823} & \mso{9837}{1887} & \mso{109}{0} & \mso{2}{1} & \mso{8}{0} & \mso{120}{0} \\
\midrule
\pfmt{p-id}    & \mso{82}{2}   & \mso{73}{0}  & \mso{5}{0}        & \mso{283}{7}    & \mso{9}{0}  & \mso{0}{0} & \mso{0}{0} & \mso{9}{0} \\
\pfmt{p-uname} & \mso{28}{0}   & \mso{19}{0}  & \mso{1}{0}        & \mso{62}{1}     & \mso{1}{0}  & \mso{5}{0} & \mso{0}{0} & \mso{6}{0} \\
\pfmt{p-cat}   & \mso{26}{1}   & \mso{30}{0}  & \mso{1}{0}        & \mso{246}{11}   & \mso{3}{0}  & \mso{1}{0} & \mso{0}{0} & \mso{4}{0} \\
\pfmt{p-ln}    & \mso{4}{0}    & \mso{36}{0}  & \mso{0}{0}        & \mso{42}{0}     & \mso{3}{0}  & \mso{0}{0} & \mso{0}{0} & \mso{3}{0} \\
\pfmt{p-date}  & \mso{160}{0}  & \mso{40}{0}  & \mso{5}{0}        & \mso{2061}{159} & \mso{5}{0}  & \mso{0}{0} & \mso{0}{0} & \mso{5}{0} \\
\pfmt{p-join}  & \mso{111}{21} & \mso{114}{0} & \mso{8}{2}        & \mso{1573}{267} & \mso{12}{0} & \mso{3}{0} & \mso{2}{1} & \mso{17}{1} \\
\pfmt{p-sort}  & \mso{116}{5}  & \mso{191}{0} & \mso{11}{1}       & \mso{3947}{167} & \mso{18}{0} & \mso{1}{0} & \mso{7}{0} & \mso{26}{0} \\
\pfmt{p-ls}    & \mso{272}{11} & \mso{216}{0} & \mso{52}{2}       & \mso{13803}{842} & \mso{25}{0} & \mso{1}{0} & \mso{3}{0} & \mso{29}{0} \\
\midrule
\pfmt{cloc}  & \mso{210}{9}   & \mso{972}{0}   & \mso{67}{5}     & \mso{5017}{456}     & \mso{21}{0} & \mso{2}{0} & \mso{13}{0} & \mso{36}{0} \\
\pfmt{ack}   & \mso{1347}{42} & \mso{867}{0}   & \mso{1962}{88}  & \mso{23127}{999}    & \mso{53}{1} & \mso{1}{0} & \mso{5}{0}  & \mso{59}{1} \\
\pfmt{grin}    & \mso{242}{28}       & \mso{331}{0}       & \mso{35}{6}        & \mso{411}{51} & \mso{9}{0} & \mso{0}{0} & \mso{9}{0} & \mso{18}{0} \\
\pfmt{pylint}  & \mso{1916}{143}       & \mso{5712}{0}       & \mso{5637}{536}   & \mso{27175}{2553} & \mso{54}{5} & \mso{1}{0} & \mso{17}{5} & \mso{72}{1} \\
\pfmt{hlint} & \mso{328}{18}  & \mso{6757}{0}  & \mso{4365}{28}  & \mso{9525}{761}     & \mso{22}{0} & \mso{1}{0} & \mso{12}{0} & \mso{35}{0} \\
\pfmt{pandoc}& \mso{653}{56}  & \mso{31635}{0} & \mso{2284}{451} & \mso{23515}{2231}   & \mso{83}{0} & \mso{7}{0} & \mso{12}{1} & \mso{102}{1} \\
\pfmt{unison}& \mso{381}{27}  & \mso{3784}{0}  & \mso{383}{38}   & \mso{4641}{341}     & \mso{36}{0} & \mso{1}{0} & \mso{13}{1} & \mso{50}{1} \\
\pfmt{bibtex2html}  & \mso{670}{118}       & \mso{1345}{0}       & \mso{369}{90}        & \mso{667}{136} & \mso{90}{0} & \mso{0}{0} & \mso{15}{0} & \mso{105}{0} \\
\pfmt{gzip}    & \mso{495}{27}   & \mso{1635}{0}  & \mso{251}{29}        & \mso{12029}{486} & \mso{27}{3} & \mso{5}{0} & \mso{11}{3} & \mso{43}{6} \\
\pfmt{httpd}    & \mso{838}{114}   & \mso{10633}{1}  & \mso{3596}{777}        & \mso{197390}{30012} & \mso{104}{3} & \mso{0}{0} & \mso{9}{6} & \mso{113}{3} \\
\midrule
\pfmt{vsftpd}& \mso{620}{38}  & \mso{2549}{0} & \mso{628}{125}   & \mso{652}{126}      & \mso{42}{0} & \mso{3}{0} & \mso{4}{0} & \mso{49}{0} \\
\pfmt{ngircd}& \mso{650}{46}  & \mso{3090}{0} & \mso{820}{88}    & \mso{1469}{125}     & \mso{34}{0} & \mso{3}{0} & \mso{9}{0} & \mso{46}{0} \\
\bottomrule
\end{tabular}
\label{tab:results_cegir}
\end{table*}

To investigate these questions, we applied iGen to the 
subject programs described in Section~\ref{sec:benchmarks}.
Table~\ref{tab:results_cegir} summarizes the results and
reports the medians across 21 runs and their 
variance\footnote{Recall from Section~\ref{sec:algorithm} that iGen uses randomness,
so different runs may produce slightly different results.} as the semi-interquartile range (SIQR). 
For each program, columns \hdrt{configs} and \hdrt{cov} show the number of 
configurations iGen created and the number of locations covered by these configurations, respectively.
The next two columns show iGen's running time in seconds: \hdrt{total} is the total time and \hdrt{search} is the time excluding the time to run the test suite.
The remaining columns list the number of inferred interactions,
divided into conjunctive (\hdrt{conj}), disjunctive (\hdrt{disj}), and
mixed interactions (\hdrt{mix}), with the cumulative sum on the right (\hdrt{total}).
The low SIQR for inferred coverage and interactions
indicate iGen produces relatively stable output across runs.

\subsection{RQ1: Correctness}
\label{sec:correctness}

\paragraph*{Exhaustive Runs}

\begin{table}[t]
\centering
\caption{Comparing iGen to exhaustive runs.}
\begin{tabular}{cc}
\begin{tabular}{lcc}
  \hdrv{prog}  & \hdrv{$\delta$ cov} &\hdrv{f-score} \\
  \pfmt{id}   & 0  &0.98 \\
  \pfmt{uname}& 0  &0.97\\
  \pfmt{cat}  & 0  &1.00 \\
  \pfmt{mv}   & 0  &0.94\\
  \pfmt{ln}   & 0  &0.99 \\
  \pfmt{date} & 0  &0.94 \\
  \pfmt{join} & -1 &0.99 \\
  \midrule
  \pfmt{vsftpd}& 0 & 0.997 \\ 
  \bottomrule
\end{tabular}
& 
\begin{tabular}{lcc}
  \hdrv{prog}    & \hdrv{$\delta$ cov} &\hdrv{f-score} \\
  \pfmt{p-id}   & 0 &0.97 \\
  \pfmt{p-uname}& 0 &1.00 \\
  \pfmt{p-cat}  & 0 &1.00 \\
  \pfmt{p-ln}   & 0 &1.00  \\
  \pfmt{p-date} &-2 &0.35 \\
  \pfmt{p-join} & 0 &0.77  \\
  \pfmt{p-sort} & 0 &1.00 \\
  \midrule               
  \pfmt{ngircd} & 0 &0.92 \\
\bottomrule
\end{tabular}
\end{tabular}
\label{tab:cmp}
\end{table}

To measure the correctness of the inferred interactions, we first
evaluated whether iGen produces the same results with its iterative
algorithm as it could produce if it used all configurations. Recall
from Section~\ref{sec:algorithm} that running iGen with all
configurations is guaranteed to find correct interactions if they match our
templates, so this essentially gives us ground truth with respect to the test suite. 

To
perform this evaluation, we selected the fourteen programs with the
smallest configuration spaces and ran one loop of iGen with
\textsf{configs} (Figure~\ref{alg:igen}) initialized to the set of all
configurations. We also used exhaustive symbolic execution information
for \pfmt{vsftpd} and \pfmt{ngircd}~\cite{reisner2010using} to
simulate iGen running with all configurations of those programs.

Table~\ref{tab:cmp} shows these comparisons.  The second column
\textsf{$\delta$ cov} reports the differences between the number of
lines covered by iGen's regular runs and those covered by the
exhaustive runs. Overall we see iGen generates interactions with
coverage very similar to the exhaustive runs. In total,
iGen missed one line in \pfmt{join} and two lines in \pfmt{p-date}.
We investigated and found these three uncovered lines are guarded by long conjunctive
interactions. For example, iGen missed \pfmt{join.c:997}, which is
guarded by an interaction with 10 conjuncts. 

Column \textsf{f-score} measures the accuracy of iGen's runs compared
to the exhaustive runs using a \emph{balanced
  f-score}~\cite{rijsbergencj}, which ranges between 0 and 1, with 1
representing perfect agreement. In more detail, the f-score is based
on comparing for every location whether settings in the interaction for that
location match between the standard and exhaustive runs. Notice this
is a very strict test. 

Table~\ref{tab:cmp} shows that iGen generates mostly the same
interactions for most programs as the exhaustive runs. We investigated
the two outliers, \pfmt{p-date} and \pfmt{p-join}, and found the low
f-scores are due to two factors. First, these programs have few
interactions (5 for \pfmt{p-date} and 17 for \pfmt{p-join}), so a few
differences cause a large score change. 
Second, iGen almost but does
not quite infer the right interactions, which still results in an
f-score penalty. For example, for \pfmt{p-date}, instead of inferring
the (correct) interaction
$G \wedge d \wedge (\neg R \vee \neg iso8601)$, iGen infers the less
precise interaction $G \wedge d$.  Similar near-matches account for
most of the f-score differences of the remaining programs, e.g., in
\pfmt{ngircd}, iGen generates several mixed interactions where it
should generate conjunctive interactions.

\paragraph*{Manual Inspection}
We also manually analyzed the program source code to make sure certain
interesting or non-obvious interactions generated by iGen are correct. We focused on coreutils because
these programs are small enough for careful manual inspection and
they allow for interesting comparisons to PPT.

For \pfmt{p-uname}, iGen discovers an interaction
$\pfmt{a}\vee \pfmt{m}$ covering the line \pfmt{uname.pl:27}, which
prints the machine name.  This interaction thus specifies (correctly)
that the machine name is printed when given either option \pfmt{a} or
\pfmt{m}. iGen also discovers other similar interactions for
\pfmt{p-uname}, including $\pfmt{a}\vee \pfmt{o}$, which prints the
operating system name, $\pfmt{a}\vee \pfmt{n}$ to print the hostname,
etc.  For the corresponding \pfmt{uname} command iGen found
similar but longer interactions such as
$(\neg \pfmt{help} \wedge \neg \pfmt{verbose}) \wedge (\pfmt{a}\vee
\pfmt{m})$,
which prints the machine name but only when \pfmt{help} and
\pfmt{verbose} are not given (note these options are not supported by PPT).
We refer to
options like \pfmt{help} as \emph{enabling options}
\cite{reisner2010using, song2014itree}, since they must be set a
certain way to achieve significant coverage. Notice also that this is
a mixed interaction, which cannot be inferred by our prior work.

Another interesting interaction iGen found is $(\neg \pfmt{A}\wedge
\neg \pfmt{t} \wedge \neg \pfmt{T}) \wedge (\pfmt{e}\vee \pfmt{v})$
covering line \pfmt{cat.c:462} of \pfmt{cat}. Reviewing the source
code, we found this line is only executed when the global variables
\pfmt{showtab} and \pfmt{shownonprinting} are false and true, respectively.
Further examination reveals that \pfmt{showtab} is false by default
but any of options \pfmt{A}, \pfmt{t}, or \pfmt{T} cause it to be true;
thus all of these options must not be given to cover the considered line.
Moreover, \pfmt{shownonprinting} is only true when either \pfmt{e},
\pfmt{t}, \pfmt{v}, or \pfmt{A} set; thus one of these options must be
given to cover the considered line. Putting this logic together and
simplifying yields the interaction iGen discovered.

We were surprised that for \pfmt{uname}, iGen generates a
purely conjunctive interaction containing the negation of all options.
Inspection of the source code reveals this is correct---one
line of code is only invoked when no options are
given. Interestingly, \pfmt{p-uname} does not have this interaction,
and indeed does not behave as \pfmt{uname} when run with no
options.

Finally, we investigated the cases of non-zero SIQRs in Table~\ref{tab:results_cegir} and found that 
most differences involve disjunctions. For example, 13 of 21 \pfmt{uname} runs found the interaction
$\pfmt{a} \vee \pfmt{s} \vee  \pfmt{n} \vee  \pfmt{r} \vee  \pfmt{v} \vee  \pfmt{m} \vee  \pfmt{p} \vee  \pfmt{i} \vee  \pfmt{o} \vee \pfmt{help}\vee  \pfmt{version}$ 
by taking a shorter conjunctive
interaction, e.g., 
$ \pfmt{i} \wedge \neg{\pfmt{a}} \wedge \neg{ \pfmt{help}} \wedge \neg{ \pfmt{version}}$; 
modifying its settings, yielding, among others, 
$\neg{\pfmt{i}} \wedge \neg{\pfmt{a}} \wedge \neg{\pfmt{help}} \wedge \neg{\pfmt{version}}$;
and randomly assigning
remaining options to get 
$\neg{\pfmt{a}}\wedge \neg{\pfmt{s}} \wedge \neg{\pfmt{n}} \wedge \neg{\pfmt{r}} \wedge \neg{\pfmt{v}} \wedge \neg{\pfmt{m}} \wedge \neg{\pfmt{p}} \wedge \neg{\pfmt{i}} \wedge \neg{\pfmt{o}} \wedge \neg{\pfmt{help}} \wedge \neg{\pfmt{version}}$,
which is then negated. 
While iGen performs random assignment for many
different short interactions, the chance of getting that exact
conjunction are still low overall---and in 8 of 21 \pfmt{uname} runs, iGen
misses that interaction. In general, non-zero SIQRs arise from lower
probability events like this.

\subsection{RQ2: Performance}
\label{rq2}

Table~\ref{tab:results_cegir} shows that for most programs, iGen's
running time is dominated by running the test suite. Thus, programs with
larger configuration spaces tend to take longer because iGen has to
create more configurations to analyze these programs.  Nonetheless,
comparing to Table~\ref{tab:stats}, we see that iGen scales well to
large programs because it only explores a small fraction of the total
possible number of configurations. Interestingly,
Table~\ref{tab:results_cegir} shows the number of explored
configurations is not directly proportional to the configuration space
size.  For example, \pfmt{ls} has eight orders of magnitude more
possible configurations than \pfmt{sort}, but iGen only explores
1.5$\times$ more configurations.

We believe that iGen represents a good trade-off between correctness
and efficiency compared to previous work.  On the one hand, iGen
generates more configurations (and hence is slower) than iTree.  For
\pfmt{vsftpd} and \pfmt{ngircd}, iGen generates 620 and 650
configurations, respectively, while iTree creates around 100
configurations each~\cite{song2014itree}. 
However, iGen generates much more
precise interaction information---iTree reports only 4 interactions
for \pfmt{vsftpd} and 3 interactions for \pfmt{ngircd}, compared to
iGen's 49 and 46 interactions, respectively.

On the other hand, iGen runs dramatically faster than the experiments
by Reisner et al.~\cite{reisner2010using}, which took two weeks to
analyze just a few programs using a specialized cluster. Moreover,
although iGen is not guaranteed to produce precise interactions, the
results discussed in Section~\ref{sec:correctness} suggest iGen is as
precise as Reisner et al.'s work in practice.

\paragraph*{Convergence}

\begin{figure}[t]
\tikzset{every picture/.style=semithick}
\begin{tikzpicture}[yscale=0.8]
\begin{axis}[
  grid=both,
  ymin=0.12,
  ymax=1.02,
  xmin=0.0,
  xmax=1.02,
  xlabel={iterations (normalized)},
  ylabel={f-score},
  legend pos=south east,
]

\addplot[mark=none, color=blue] coordinates { (0.0322580645161,0.242735774969658) (0.0645161290323,0.345947396672034) (0.0967741935484,0.381604171459244) (0.129032258065,0.585364388730586) (0.161290322581,0.686729074555593) (0.193548387097,0.710039934052317) (0.225806451613,0.890973909201664) (0.258064516129,0.946451917657523) (0.290322580645,0.960177355075377) (0.322580645161,0.964914242584004) (0.354838709677,0.974224769660057) (0.387096774194,0.980042717373002) (0.41935483871,0.982323661390663) (0.451612903226,0.982543248566772) (0.483870967742,0.982543248566772) (0.516129032258,0.982543248566772) (0.548387096774,0.982543248566772) (0.58064516129,0.982543248566772) (0.612903225806,0.982543248566772) (0.645161290323,0.982543248566772) (0.677419354839,0.982543248566772) (0.709677419355,0.982543248566772) (0.741935483871,0.982543248566772) (0.774193548387,0.982543248566772) (0.806451612903,0.982543248566772) (0.838709677419,0.982543248566772) (0.870967741935,0.982543248566772) (0.903225806452,0.982543248566772) (0.935483870968,0.982543248566772) (0.967741935484,0.982543248566772) (1.0,0.982543248566772) };
\addplot[mark=none, color=red] coordinates { (0.0384615384615,0.375973303670745) (0.0769230769231,0.788414142075072) (0.115384615385,0.9405278422319) (0.153846153846,0.954910379537606) (0.192307692308,0.963089564297435) (0.230769230769,0.958750853666108) (0.269230769231,0.953129210123355) (0.307692307692,0.953498317967982) (0.346153846154,0.952834485690699) (0.384615384615,0.952834485690699) (0.423076923077,0.972178703778478) (0.461538461538,0.972507111003437) (0.5,0.972507111003437) (0.538461538462,0.972507111003437) (0.576923076923,0.972507111003437) (0.615384615385,0.972507111003437) (0.653846153846,0.972507111003437) (0.692307692308,0.972507111003437) (0.730769230769,0.972507111003437) (0.769230769231,0.972507111003437) (0.807692307692,0.972507111003437) (0.846153846154,0.972507111003437) (0.884615384615,0.972507111003437) (0.923076923077,0.961012858129873) (0.961538461538,0.961012858129873) (1.0,0.972507111003437) };
\addplot[mark=none, color=green] coordinates { (0.0454545454545,0.223068050749712) (0.0909090909091,0.5986475754889) (0.136363636364,0.760773289107038) (0.181818181818,0.817213195578101) (0.227272727273,0.826753660613335) (0.272727272727,0.846300235980643) (0.318181818182,0.981066376645211) (0.363636363636,0.997382270708842) (0.409090909091,1) (0.454545454545,1) (0.5,1) (0.545454545455,1) (0.590909090909,1) (0.636363636364,1) (0.681818181818,1) (0.727272727273,1) (0.772727272727,1) (0.818181818182,1) (0.863636363636,1) (0.909090909091,1) (0.954545454545,1) (1.0,1) };
\addplot[mark=none, color=black] coordinates { (0.05,0.590647070647785) (0.1,0.758827530434177) (0.15,0.884775756027486) (0.2,0.917722526912794) (0.25,0.935889725869206) (0.3,0.936024797515905) (0.35,0.935906594206213) (0.4,0.936153996482314) (0.45,0.936153996482314) (0.5,0.935906594206213) (0.55,0.936153996482314) (0.6,0.936153996482314) (0.65,0.936153996482314) (0.7,0.936153996482314) (0.75,0.936153996482314) (0.8,0.936153996482314) (0.85,0.936153996482314) (0.9,0.937926645840565) (0.95,0.937040321161439) (1.0,0.936153996482314) };
\addplot[mark=none, color=cyan] coordinates { (0.0227272727273,0.56783941518807) (0.0454545454545,0.762696963304935) (0.0681818181818,0.827172077121196) (0.0909090909091,0.898658313436743) (0.113636363636,0.92254920066373) (0.136363636364,0.94789844042291) (0.159090909091,0.967245900073149) (0.181818181818,0.978267174646366) (0.204545454545,0.980689890032044) (0.227272727273,0.982132942388975) (0.25,0.982132942388975) (0.272727272727,0.98354940307682) (0.295454545455,0.98354940307682) (0.318181818182,0.98951967791665) (0.340909090909,0.989191783352299) (0.363636363636,0.989191783352299) (0.386363636364,0.98951967791665) (0.409090909091,0.98951967791665) (0.431818181818,0.989580347761514) (0.454545454545,0.989580347761514) (0.477272727273,0.989580347761514) (0.5,0.989580347761514) (0.522727272727,0.989580347761514) (0.545454545455,0.990079621545101) (0.568181818182,0.990079621545101) (0.590909090909,0.990586784844662) (0.613636363636,0.990271562709879) (0.636363636364,0.990271562709879) (0.659090909091,0.990586784844662) (0.681818181818,0.990854744140399) (0.704545454545,0.99104507141406) (0.727272727273,0.992316543915691) (0.75,0.992316543915691) (0.772727272727,0.992316543915691) (0.795454545455,0.991581172762691) (0.818181818182,0.992776347351511) (0.840909090909,0.992546445633601) (0.863636363636,0.992316543915691) (0.886363636364,0.99294836359513) (0.909090909091,0.99294836359513) (0.931818181818,0.99294836359513) (0.954545454545,0.995265839220088) (0.977272727273,0.992891881780771) (1.0,0.992891881780771) };
\addplot[mark=none, magenta] coordinates { (0.0227272727273,0.31240186949299) (0.0454545454545,0.434068777252956) (0.0681818181818,0.530031147475436) (0.0909090909091,0.646160951872639) (0.113636363636,0.727826210576408) (0.136363636364,0.756276848967848) (0.159090909091,0.787768680944234) (0.181818181818,0.798418817121584) (0.204545454545,0.800352193722946) (0.227272727273,0.813838700489391) (0.25,0.822740929613561) (0.272727272727,0.840844425388628) (0.295454545455,0.845332373077495) (0.318181818182,0.85458442666679) (0.340909090909,0.875526798149011) (0.363636363636,0.877291223031235) (0.386363636364,0.886011293402562) (0.409090909091,0.892495723597302) (0.431818181818,0.892313250617447) (0.454545454545,0.892123932261108) (0.477272727273,0.891859020728903) (0.5,0.892130442618077) (0.522727272727,0.890778537111491) (0.545454545455,0.896153655524201) (0.568181818182,0.896153655524201) (0.590909090909,0.906489230699979) (0.613636363636,0.905090074331092) (0.636363636364,0.905090074331092) (0.659090909091,0.921854107859159) (0.681818181818,0.921854107859159) (0.704545454545,0.913724401629538) (0.727272727273,0.921854107859159) (0.75,0.921854107859159) (0.772727272727,0.925207163049149) (0.795454545455,0.93017179670723) (0.818181818182,0.956048675733715) (0.840909090909,0.956048675733715) (0.863636363636,0.956048675733715) (0.886363636364,0.956418515867335) (0.909090909091,0.956484674952821) (0.931818181818,0.971019469843061) (0.954545454545,0.956233818049535) (0.977272727273,0.956484674952821) (1.0,0.956484674952821) };
\addplot[mark=none,color=brown] coordinates { (0.0161290322581,0.543120482861781) (0.0322580645161,0.681432805172143) (0.0483870967742,0.728348561565496) (0.0645161290323,0.86038217114995) (0.0806451612903,0.906602687878803) (0.0967741935484,0.921039775498613) (0.112903225806,0.932359870856562) (0.129032258065,0.942485074862401) (0.145161290323,0.951982223894213) (0.161290322581,0.961278285245823) (0.177419354839,0.968689255446098) (0.193548387097,0.972103893467469) (0.209677419355,0.980585569352676) (0.225806451613,0.98108882756404) (0.241935483871,0.98108882756404) (0.258064516129,0.981562606849862) (0.274193548387,0.982858751846033) (0.290322580645,0.982858751846033) (0.306451612903,0.982858751846033) (0.322580645161,0.982858751846033) (0.338709677419,0.983577720154556) (0.354838709677,0.984976015488562) (0.370967741935,0.984802359846541) (0.387096774194,0.984730901828127) (0.403225806452,0.984980401697084) (0.41935483871,0.98596990690508) (0.435483870968,0.98596990690508) (0.451612903226,0.98596990690508) (0.467741935484,0.985946412296582) (0.483870967742,0.985946412296582) (0.5,0.985946412296582) (0.516129032258,0.985946412296582) (0.532258064516,0.985951257200135) (0.548387096774,0.98621409559427) (0.564516129032,0.98657577717395) (0.58064516129,0.986836798485085) (0.596774193548,0.986836798485085) (0.612903225806,0.987179688640356) (0.629032258065,0.987179688640356) (0.645161290323,0.987296315997311) (0.661290322581,0.987296315997311) (0.677419354839,0.987775651264396) (0.693548387097,0.987775651264396) (0.709677419355,0.987775651264396) (0.725806451613,0.98819181142006) (0.741935483871,0.98834312658662) (0.758064516129,0.987775651264396) (0.774193548387,0.987613287694524) (0.790322580645,0.987775651264396) (0.806451612903,0.987983731342228) (0.822580645161,0.988580487137297) (0.838709677419,0.988666532521413) (0.854838709677,0.988580487137297) (0.870967741935,0.988666532521413) (0.887096774194,0.989343740916507) (0.903225806452,0.989343740916507) (0.91935483871,0.989343740916507) (0.935483870968,0.989343740916507) (0.951612903226,0.988666532521413) (0.967741935484,0.986665863378802) (0.983870967742,0.986665863378802) (1.0,0.990020949311602) };

\legend{id, uname, cat, mv, ln, date, join}
\end{axis}
\end{tikzpicture}
\caption{Progress of iGen on generating interactions for GNU coreutils programs.}
\label{fig:evolv1}
\end{figure}
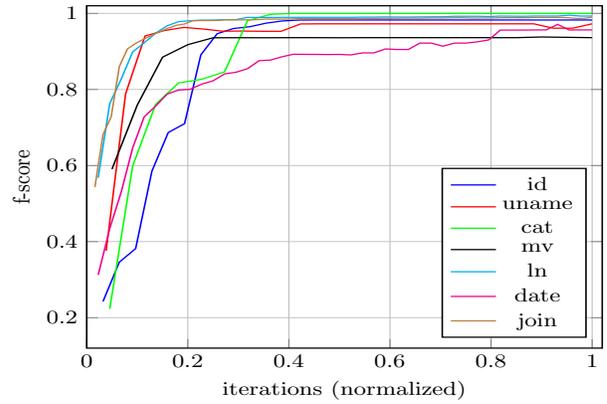

Figure~\ref{fig:evolv1} shows how iGen converges to its final results
on a subset of coreutils. The $x$-axis is the iteration count
(normalized such that 1 represents all iterations for that particular
program), and the $y$-axis is the f-score compared to the exhaustive
runs. These results show that iGen converges fairly quickly. After
approximately 20\% of the iterations, iGen's f-score has reached 0.8
or more. This suggests an iGen user could potential cut off iteration
early and still achieve reasonable results.

\begin{table}[t]
\centering
\caption{Comparing random search to exhaustive runs.}
\begin{tabular}{ll}
\begin{tabular}{lrr}
  \hdrv{prog}  & \hdrv{$\delta$ cov} &\hdrv{f-score} \\
  \pfmt{id}   & -3  &0.85 \\
  \pfmt{uname}& -1  &0.83\\
  \pfmt{cat}  & -1  &0.93 \\
  \pfmt{mv}   &  0  &0.58\\
  \pfmt{ln}   & -4  &0.76 \\
  \pfmt{date} & -4  &0.51 \\
  \pfmt{join} &-17  &0.82 \\
  \midrule
  \pfmt{vsftpd}& -356 & 0.58 \\ 
  \bottomrule
\end{tabular}
& 
\begin{tabular}{lrr}
  \hdrv{prog}    & \hdrv{$\delta$ cov} &\hdrv{f-score} \\
  \pfmt{p-id}   & -2 &0.88 \\
  \pfmt{p-uname}&  0 &0.94 \\
  \pfmt{p-cat}  & -1 &0.93 \\
  \pfmt{p-ln}   &  0 &1.00 \\
  \pfmt{p-date} & -2 &0.33 \\
  \pfmt{p-join} & -6 &0.70 \\
  \pfmt{p-sort} &  -2 &0.90 \\
  \midrule               
  \pfmt{ngircd} & -1289 &0.27 \\
\bottomrule
\end{tabular}
\end{tabular}
\label{tab:cmp_rand}
\end{table}

iGen's convergence relies critically on \textsf{genNewConfigs}
(Figure~\ref{alg:igen} line~\ref{line:gennewconfigs}) generating useful
configurations. Table~\ref{tab:cmp_rand} demonstrates this by showing
iGen runs on randomly selected configurations. More specifically, for
each program in Table~\ref{tab:cmp}, we generated the same number of
random configurations as the number of configurations iGen used (Table~\ref{tab:results_cegir}).
We also include the default configuration, if it exists.
We next ran the iGen main loop once on these configurations to compute the results,
and then compared coverage information and f-score to the exhaustive
runs.

Comparing to Table~\ref{tab:cmp}, we see iGen
generates much more precise interactions and has better coverage than
random search. The differences are most significant 
for larger programs, e.g., the most extreme case is
\textsf{ngircd}, where random search has f-score 0.27 and covers
1289 fewer lines than exhaustive run, while iGen has f-score
0.92 and has the same coverage as the exhaustive run.

\subsection{RQ3: Analysis}

We analyzed iGen's results in detail to learn interesting
properties of the subject programs. In the following,
an interaction's \emph{length} is the number of options it
contains.\footnote{It is more typical to refer to this as the \emph{strength}
  of an interaction. However, in our setting, longer conjunctive
  formulae are logically stronger than shorter conjunctive formulae,
  but longer disjunctive formulae are logically weaker than shorter
   disjunctive formulae. Hence we use \emph{length} to avoid confusion.}

\begin{table*}[t]
\small
\centering
\caption{Number of interactions and covered lines per interaction length. Results are medians of 21 runs.}
\begin{tabular}{l@{\;}|r@{\ }r|r@{\ }r|r@{\ }r|r@{\ }r|r@{\ }r|r@{\ }r|r@{\ }r|r@{\ }r|r@{\ }r|r@{\ }r|r@{\ }r|r}
\hdrv{prog}&\multicolumn{2}{c}{\hdrv 0}&\multicolumn{2}{c}{\hdrv 1}&\multicolumn{2}{c}{\hdrv 2}&\multicolumn{2}{c}{\hdrv 3}&\multicolumn{2}{c}{\hdrv 4}&\multicolumn{2}{c}{\hdrv 5}&\multicolumn{2}{c}{\hdrv 6}&\multicolumn{2}{c}{\hdrv 7}&\multicolumn{2}{c}{\hdrv 8}&\multicolumn{2}{c}{\hdrv 9}&\multicolumn{2}{c}{\hdrv 10+}&{\hdrv max}\\
\pfmt{id}      & 1 & 15    &  2 &  3    &  7 & 29    &  2 & 29   &  2 & 10  &  1 & 1    &  5 & 14 &  1 & 2  &  1 & 1  &  2 & 33 &  1 & 1  &  10\\
\pfmt{uname}   & 1 & 10    & 10 &  32   &  2 & 32    &  - & -    &  9 & 11  &  - & -    &  - & -  &  - & -  &  - & -  &  - & -  &  2 & 2  &  11\\
\pfmt{cat}     & 1 & 16    & 11 &  42   &  2 & 35    &  - & -    &  2 & 13  &  1 & 1    &  1 & 22 &  2 & 3  &  1 & 1  &  - & -  &  4 & 71 &  12\\
\pfmt{mv}      & 1 & 51    &  9 &  36   &  3 & 53    &  3 & 11   &  1 & 1   &  1 & 21   &  - & -  &  - & -  &  - & -  &  - & -  &  - & -  &   5\\
\pfmt{ln}      & 1 & 12    & 10 &  44   &  3 & 38    &  1 & 1    &  4 & 54  &  3 & 7    &  2 & 3  &  2 & 3  &  - & -  &  - & -  &  - & -  &   7\\
\pfmt{date}    & 1 & 14    &  3 &  9    &  3 & 89    &  3 & 9    &  - & -   &  - & -    &  4 & 6  &  - & -  &  - & -  &  - & -  &  - & -  &   6\\
\pfmt{join}    & 1 & 21    & 10 &  66   &  7 & 197   & 10 & 51   &  6 & 28  &  5 & 10   &  1 & 6  &  1 & 3  &  - & -  &  - & -  &  - & -  &   7\\
\pfmt{sort}    & 1 & 82    & 11 &  25   &  4 & 183   & 18 & 70   &  6 & 82  &  7 & 344  & 13 & 92 &  4 & 48 &  3 & 11 &  2 & 3  & 21 & 87 &  15\\
\pfmt{ls}      & 1 & 51    & 46 &  160  &  3 & 187   & 11 & 317  & 22 & 164 & 13 & 62   &  9 & 25 &  5 & 29 &  2 & 8  &  1 & -  & 13 & 26 &  47\\
\midrule
\pfmt{p-id}    & 1 & 7     &  1 &  11   &  - & -     &  - & -    &  - & -   &  5 & 53   &  2 & 2  &  - & -  &  - & -  &  - & -  &  - & -  &  6 \\
\pfmt{p-uname} & 1 & 14    &  - &  -    &  5 & 5     &  - & -    &  - & -   &  - & -    &  - & -  &  - & -  &  - & -  &  - & -  &  - & -  &  2 \\
\pfmt{p-cat}   & 1 & 23    &  2 &  4    &  - & -     &  1 & 3    &  - & -   &  - & -    &  - & -  &  - & -  &  - & -  &  - & -  &  - & -  &  3 \\
\pfmt{p-ln}    & 1 & 34    &  2 &  2    &  - & -     &  - & -    &  - & -   &  - & -    &  - & -  &  - & -  &  - & -  &  - & -  &  - & -  &  1 \\
\pfmt{p-date}  & 1 & 8     &  4 &  32   &  - & -     &  - & -    &  - & -   &  - & -    &  - & -  &  - & -  &  - & -  &  - & -  &  - & -  &  1 \\
\pfmt{p-join}  & 1 & 59    & 10 &  32   &  2 & 6     &  - & -    &  - & -   &  1 & 7    &  - & -  &  - & -  &  - & -  &  - & -  &  1 & 4  & 10 \\
\pfmt{p-sort}  & 1 & 42    &  6 &  43   &  6 & 34    &  2 & 8    &  3 & 49  &  5 & 9    &  2 & 4  &  1 & 2  &  - & -  &  - & -  &  - & -  &  7 \\
\pfmt{p-ls}    & 1 & 124   &  7 &  49   &  9 & 25    &  2 & 4    &  3 & 5   &  5 & 7    &  2 & 2  &  - & -  &  - & -  &  - & -  &  - & -  &  6 \\
\midrule
\pfmt{cloc}    & 1 & 190   &  4 &  567  & 11 & 124   &  7 & 50   &  8 & 29  &  1 & 1    &  4 & 11 &  - & -  &  - & -  &  - & -  &  - & -  &  6 \\
\pfmt{ack}     & 1 & 105   &  1 &  2    &  2 & 7     &  2 & 17   &  2 & 14  &  2 & 57   &  1 & 36 &  2 & 48 &  4 & 52 &  6 & 57 & 34 & 464& 15 \\
\pfmt{grin}    &	1	&	105	&	2	&	73	&	5	&	112	&	0	&	0	&	3	&	11	&	3	&	23	&	3	&	5	&	1	&	2	&	-	&	-	&	-	&	-	&	-	&	- & 7 \\
\pfmt{pylint}    &	1	&	2062	&	1	&	4	&	3	&	44	&	5	&	39	&	2	&	101	&	6	&	348	&	15	&	1602	&	18	&	1436	&	13	&	62	&	6	&	12	& -	&	-	&	9 \\
\pfmt{hlint}   & 1 & 288   &  1 &  13   &  2 & 2270  &  4 & 150  &  8 & 324 & 10 & 2978 &  3 & 61 &  3 & 8  &  2 & 641&  1 & 24 &  - & -  &  9 \\
\pfmt{pandoc}  & 1 & \num{27788} & 60 &  2674 & 25 & 1065  & 10 & 86   &  2 & 9   &  1 & 2    &  1 & 1  &  - & -  &  - & -  &  1 & 1  &  - & -  &  5 \\
\pfmt{unison}  & 1 & 983   &  3 &  1970 & 16 & 579   & 12 & 137  & 14 & 102 &  3 & 10   &  1 & 1  &  - & -  &  - & -  &  - & -  &  - & -  &  6 \\
\pfmt{bibtex2html}  &	1	&	372	&	35	&	419	&	6	&	149	&	19	&	143	&	16	&	150	&	22	&	90	&	5	&	19	&	-	&	-	&	-	&	-	&	-	&	-	&	-	&	-	&	6 \\
\pfmt{gzip}    &	1	&	103	&	2	& 9	&	4	& 182	&	6	& 370	&	5	& 206	&	4	& 86	&	4	& 12	&	3	& 19	&	2	& 28	&	1	& 1	&	7	& 47 & 17 \\
\pfmt{httpd}	&	1	&	104	&	1	&	708	&	39	&	5524	&	46	&	3765	&	13	&	370	&	4	&	40	&	3	&	90	&	1	&	4	&	-	&	-	&	-	&	-	&	50	&	1	&	50 \\
\midrule
\pfmt{vsftpd}  & 1 & 336   &  4 &  101  &  6 & 170   &  4 & 1373 & 18 & 410 &  8 & 114  &  6 & 35 &  2 & 10 &  - & -  &  - & -  &  - & -  &  7 \\
\pfmt{ngircd}  & 1 & 748   &  3 &  460  &  4 & 525   & 16 & 827  & 14 & 457 &  6 & 68   &  1 & 2  &  - & -  &  - & -  &  - & -  &  - & -  &  6 \\
\bottomrule
\end{tabular}
\label{tab:results_count}
\end{table*}

\paragraph*{Disjunctive and Mixed Interactions}

We observe that many programs require non-conjunctive interactions.
Table~\ref{tab:results_cegir} shows that approximately 20\% of the
inferred interactions are disjunctions or mixed
interactions. 
Furthermore, the analysis in
Section~\ref{sec:correctness} shows that mixed interactions, though
rare, do exist in real-world software.  Thus, these interactions,
which were omitted from prior work~\cite{reisner2010using,
  song:icse12, song2014itree}, are important.

\paragraph*{Interaction Length and Coverage}
In prior work we observed that the total
number of interactions found is far fewer than the number of possible
interactions~\cite{song2014itree,song:icse12}.
We observe the same trend in Table~\ref{tab:results_cegir}.
For example, \pfmt{p-cat} has $7$ binary options, yielding
$128$ possible configurations and at least $4373$ possible
interactions. 
However, iGen finds four
interactions, which is less than $0.1\%$ of $4373$. 
The same trend can be seen throughout the table.

We also observed in prior work that longer conjunctive
interactions tend to contain shorter conjunctive
interactions, e.g., if $a~ \wedge ~b~ \wedge ~c~ \wedge ~d~ \wedge ~e$ is an
interaction, it is likely that a shorter formula like $a~ \wedge ~b$ is
an interaction~\cite{reisner2010using}.
We manually examined iGen's interactions and found that this pattern also holds.
For most programs, conjunctive interactions of length at least three
include a shorter interaction. This is likely due to nested
guards, e.g., such as the interaction on $L5$ of Figure~\ref{fig1}, but
with a non-disjunctive inner condition.

Table~\ref{tab:results_count} looks at the interactions in more detail 
by showing the number of interactions iGen
infers and the covered lines at each interaction length. The
last column (\hdrt{max}) reports the length of the longest
interaction.
We observe that the maximum interaction length for most programs is 
significantly smaller than the number of configuration options. 
However, there are five programs---\pfmt{id}, \pfmt{uname}, \pfmt{cat},
\pfmt{p-join}, and \pfmt{httpd}---that have interactions that include all
options. (We discussed the \pfmt{uname} case in Section~\ref{sec:correctness}.)

Although some non-trivial coverage is achieved by large interactions 
(e.g., conjunctions of all options), 87\% of the coverage is obtained 
by interactions of length less than three.
These observations are similar to our prior 
work~\cite{reisner2010using,song2014itree,song:icse12}.

One exceptional case is \pfmt{ack}, in which most coverage is achieved
by large interactions. Investigating further, we found \pfmt{ack} has
many options that must be disabled for most of its functionality
to be exercised.

\paragraph*{Enabling Options}

In addition to the enabling option \pfmt{help} mentioned for
\pfmt{uname}, and the enabling options just mentioned for \pfmt{ack}, we found
many other examples of enabling options, including the following.
For \pfmt{id}, option \pfmt{Z} must be disabled for most coverage
(because that option is only applicable to a secured Linux kernel).
For \pfmt{httpd}, option \pfmt{--enable-http} must be enabled for
almost all coverage, and \protect{\pfmt{--enable-so}} (which allows for shared
modules) must be enabled for a majority of the coverage.
For \pfmt{vsftpd}, disabling \pfmt{ssl} and \pfmt{local}, and
enabling \pfmt{anon}, are important to coverage.
Finally, for \pfmt{ngircd}, options \pfmt{ListenIPv4} and
\pfmt{Conf\_PredefChannelsOnly} are important to coverage.

Notice these results depend on the test suite and
environment. For example, an enabled option \pfmt{Z} for \pfmt{id} could be
useful on a secured kernel, and \pfmt{ssl} for \pfmt{vsftpd}
can be enabled when running with SSL certificates.
This shows how iGen naturally adapts the 
inferred interactions to the current setting.

\paragraph*{Minimal Covering Configurations}
Finally, we used inferred interactions to compute a minimal set of
configurations that achieve the full coverage found by iGen.
To do so, we used the following greedy algorithm.  Given a set of
interactions, we first remove any interactions implied by others. For
example, if $x~\wedge~ y$ and $x$ are interactions, we remove $x$,
because a configuration satisfying $x\wedge y$ will also satisfy
$x$. Next, we randomly conjoin interactions whose conjunction
is satisfiable, e.g., we combine $x~\wedge~ y$ and $z ~\vee~ w$ to yield
$x ~\wedge ~y \wedge ~(z ~\vee~ w)$. 
This operation is greedy because it tries to combine as many 
compatible interactions as possible.
Finally, we generate configurations
that are compatible with the combined interactions, thus producing a
small set of configurations that covers the same locations as the
interactions. Note that this algorithm may not produce an optimal
result, but in practice it is effective.

\begin{table}[t]
\centering
\caption{Using iGen's interactions to compute minimal covering configurations.}
\small
\begin{tabular}{cc}
  \begin{tabular}[t]{@{}l@{}r@{\ }r@{}}
  \cellcolor{black!30}& \cellcolor{black!30}&\hdrr{min}\\
  \hdrv{prog} & \hdrr{cspace} &\hdrr{cspace} \\
  \pfmt{id}   & 1024          & 10\\
  \pfmt{uname}& 2048          & 4\\
  \pfmt{cat}  & 4096          & 6\\
  \pfmt{mv}   & 5120          & 4\\
  \pfmt{ln}   & \num{10240}   & 7 \\
  \pfmt{date} & \num{17280}   & 7 \\
  \pfmt{join} & \num{18432}   & 7 \\
  \pfmt{sort} & \num{6291456} & 17 \\
  \pfmt{ls}   & $3.5$$\times$$10^{14}$ & 15 \\
  \midrule
  \pfmt{p-id}   & 256  & 7\\
  \pfmt{p-uname} & 64   & 1\\
  \pfmt{p-cat}  & 128  & 1\\
  \pfmt{p-ln}   & 4    & 1 \\
  \pfmt{p-date} & 3360 & 3\\
  \pfmt{p-join} & 4608 & 4 \\
    \bottomrule
\end{tabular}& 
  \begin{tabular}[t]{@{}lr@{\ }r@{}}
  \cellcolor{black!30}& \cellcolor{black!30} &\hdrr{min}\\
  \hdrv{prog} & \hdrr{cspace} &\hdrr{cspace} \\
  \pfmt{p-sort} & 2048 & 6\\
  \pfmt{p-ls}   & $6.7$$\times$$10^7$ & 10 \\
  \pfmt{cloc}    & \num{524288} & 6 \\
  \pfmt{ack}     & $4.3$$\times$$10^9$  & 13 \\
  \pfmt{grin}    & \num{2097152} & 6 \\
  \pfmt{pylint}  & $5.8$$\times$$10^{10}$ & 11 \\
  \pfmt{hlint}   & \num{8192} & 5  \\
  \pfmt{pandoc}  & $4.0$$\times$$10^9$ & 5 \\
  \pfmt{unison}  & \num{393216} & 7 \\
  \pfmt{bibtex2html} &  $1.2$$\times$$10^9$ &8\\
  \pfmt{gzip}    & \num{131072} & 10 \\
  \pfmt{httpd}   & $1.1$$\times$$10^{15}$ & 5 \\   
  \midrule                  
  \pfmt{vsftpd} & $2.1$$\times$$10^9$ & 6 \\  
  \pfmt{ngircd} & \num{29764} & 6 \\
    \bottomrule
\end{tabular}
\end{tabular}
\label{tab:minconfigs}
\end{table}

Table~\ref{tab:minconfigs} summarizes the results. For each program we
list the full configuration space (from Table~\ref{tab:stats}) and the
size of the minimal configuration set computed with our algorithm 
(again the coverage of these sets are similar as those as shown 
in Table~\ref{tab:results_cegir}). 
Our results show that the minimal configuration set is dramatically
smaller than the full configuration space. In prior work, we found
similar results: We constructed a minimal line covering set of 5
configurations for \pfmt{vsftpd} and 7 for
\pfmt{ngricd}~\cite{reisner2010using}. We believe the small
differences between those results and Table~\ref{tab:minconfigs} are
due to iGen supporting richer interactions and randomness in the
greedy algorithm.

\subsection{Threats to Validity}

Like any empirical study, our conclusions are limited
by potential threats to validity. As a result our findings may not generalize
in certain ways or to certain systems.

In this work we examined several subject systems,
covering a range of different sizes, from 25 to 238k lines of code,
written in five different languages. Although each of these systems
is realistic and widely used, the whole set of systems represents only
a sample of all possible software systems. In addition, we focused on
subsets of configuration options; the number of options was
substantial, but we did not include every possible option, as
discussed earlier.

Another potential threat is that iGen relies on running test suites
to draw its conclusions. The test sets we used have reasonable,
but not complete, coverage. Individually, the test cases tend to focus
on specific functionality, rather than combining multiple activities
in a single test case. In that sense they are more like a typical
regression suite than a customer acceptance suite. Systems whose test
suites are less (or more) complete could have different results.

Finally, iGen misses interactions that are not in one of the
forms discussed in Section~\ref{sec:algorithm}.
However, from Table~\ref{tab:results_cegir} we see that interactions
with disjunctions are much less common than those solely composed
of conjunctions. Hence we speculate even more complex interaction
forms are uncommon.

We intend to explore each of these issues in future studies.

\section{Related Work}\label{sec:related}

There are several threads of related work.

\paragraph*{Interaction Discovery}

As discussed earlier, Reisner et al. \cite{reisner2010using} use the
symbolic executor Otter to fully explore the configuration space of a
software system and extract interactions from the resulting
information. In that work, interactions were limited to conjunctions,
while iGen supports much richer interaction templates. Moreover, while
symbolic execution is powerful, it suffers major limitations. First,
experience shows it has limited scalability. For example, the Otter
experiments required several days on a large cluster and analyzed only
a few programs. In contrast, Table~\ref{tab:results_cegir} shows that
iGen is much more efficient. Second, symbolic executors are
language-specific, e.g., Otter could not be applied to C++ or Haskell.
In contrast, the only language-specific tools iGen relies on are
code coverage tools, which are easy to use and widely available for
many languages. 
Finally, symbolic execution is very hard to apply to
programs that use frameworks, libraries, and/or native
code. 
Typically symbolic execution users must replace these parts of a
system with painstakingly developed ``stub'' code that implements the
functionality in a more symbolic executor-friendly way. Needless to
say this process is time-consuming, error-prone, and hard to maintain
as systems evolve.

As also discussed earlier, to address some limitations of Otter we
developed iTree~\cite{song2014itree}, which uses dynamic analysis and
machine learning techniques to generate a set of configurations that
achieve high coverage. iTree works by constructing an ``interaction
tree,'' where each node of the tree is a formula, and conjoining the
formulae on a path from the root to a node yields a potential
interaction. However, while iTree does infer some interactions, its
main goal is to achieve high coverage. As a result, as we saw in
Section~\ref{rq2}, iTree does not actually infer very many, or very useful,
interactions. Moreover, as
with Otter, iTree's interactions are limited to conjunctions, although
iTree has some support for membership-like constraints to improve
efficiency. 

\paragraph*{Feature Interactions and Presence Conditions}
The concepts of \emph{feature interactions} and \emph{presence conditions} are similar to our use of \emph{interactions}.
Th\"{u}m et al~\cite{thum2012analysis} classify problems in software product line research and surveys existing static analysis to solve them.
Our use of interactions belongs to the feature-based classification, and we propose a new dynamic analysis to generate them.  
 Apel et al~\cite{apel2013exploring} study the number of feature interactions in a system and their effects, including bug triggering, power consumption, etc. 
Our work complements these results by studying interactions that
affect line or expression coverage. 
Lillack et al~\cite{lillack2014tracking} use (language-specific) taint analysis to find interactions in Android applications. 
In contrast, iGen uses a dynamic analysis that is language agnostic (but potentially unsound). 
Nadi et al~\cite{nadi2015configuration} and von Rhein et al~\cite{von2015presence} propose tools that work with presence conditions that are already provided. 
In contrast, iGen discovers interactions. 
Czarnecki and Pietroszek~\cite{czarnecki2006verifying} check for well-formness errors in UML featured-based model templates using an SAT solver. 
We intend to explore SMT-based techniques to verify correctness of iGen's inferred interactions.

\paragraph*{Combinatorial Interaction Testing}

Many researchers have explored combinatorial interaction testing
(CIT)~\cite{tai2002test, cohen1996combinatorial, nie2011survey}, a
family of techniques for testing a program under a systematically
generated set of configurations.
One particularly popular approach is called $t$-way covering arrays,
which, given an interaction \emph{strength}~$t$, generate a set of
configurations containing all $t$-way combinations of option
settings at least once. Over last 30 years, many studies have focused on improving 
the speed, quality and flexibility of covering arrays \cite{bryce2006prioritized,
  yuan2011gui, cohen2003constructing, demiroz2012cost,
  yilmaz2006covering}.  However, as pointed out in Fouche
et al.~\cite{Fouche09-IncrementalCA}, because developers must choose
$t$ a priori and because generating covering arrays quite expensive,
developers will often set $t$ to be small, causing higher strength interactions to be ignored.

\paragraph*{Invariant Generation}

Interactions can also be considered \emph{invariants} that hold at
particular locations, i.e., specific option must have specific
settings whenever execution reaches that location. Thus, iGen can be
seen as a likely invariant generator that works for one particular
class of invariants, those restricted to configuration options and with
specific forms (quantifier-free expressions involving only equalities
and certain conjunctions and disjunctions).
 
Other researchers have proposed general-purpose invariant
generators. Daikon~\cite{daikon} is a well-known dynamic invariant
generation system that works by hypothesizing many potential
invariants and then using run-time monitoring to eliminate those that
do not hold. Daikon includes a large list of invariant templates. DIG
is a more recent invariant generator that supports more expressive
invariants including nonlinear arithmetic, disjunctive polynomials,
and relations among arrays~\cite{nguyen2014dig}. 
DySy is another
invariant generator that uses symbolic execution for invariant
inference~\cite{csallner2008dysy}.

iGen differs from Daikon, DIG, and DySy in three main ways. First,
iGen can compute potentially long disjunctive invariants very
efficiently (via pointwise union and negation). 
In contrast, Daikon requires users to provide ``splitting''
conditions~\cite{ernstcollections} to find disjunctive invariants
such as ``$\mathsf{if}\;c \; \mathsf{then}\; a\; \mathsf{else}\; b$'',
and DIG uses complex and expensive-to-compute convex hulls to
represent disjunctions. 
DySy is bounded by the limitations of symbolic execution
---the language-specificity and the difficult of analyzing frameworks, 
libraries, and native code.
Second, Daikon, DIG, and DySy do not attempt to 
search for more executions to refine invariants. Finally, iGen tries to find 
interactions for every location, while previous work only 
considers specific locations (e.g., loops and function entrance and exit) 
to reduce run-time overhead.

\section{Conclusion}

We presented iGen, a new, lightweight approach to infer interactions,
which are formulae that describe the configurations covering a
location. iGen discovers interactions by running the subject program
under a set of configurations to determine coverage information;
computing the pointwise union of various sets of covering and
non-covering configurations; and then combining the resulting formulae
to produce an interaction. iGen repeats this process, iteratively
refining the set of configurations, until no new coverage is
achieved and no new interactions are produced. We applied iGen
to 29 programs written in five different languages and demonstrated
that iGen infers precise interactions; it does so using a small
fraction of the number of possible configurations; iGen
confirms several observations made by prior work; and the disjunctive
and mixed interactions inferred by iGen occur with nontrivial
frequency.
We believe iGen takes an important step forward in the practical
understanding of configurable systems.

\section{ACKNOWLEDGMENTS}
We thank the anonymous reviewers for their many helpful comments.
This research was supported in part by NSF CCF-1116740, CCF-1139021,
and CCF-1319666.

\balance
\bibliographystyle{abbrv}
\bibliography{vu_bibs,vu_bibs1}  
\end{document}